\documentclass[aps,pra,twocolumn,groupedaddress,floatfix,showpacs]{revtex4}

\usepackage{graphicx}
\usepackage[latin1]{inputenc}
\usepackage{amsmath}
\usepackage{amssymb}
\usepackage{color}

\newcommand{\nc}{\newcommand}
\nc{\bra}[1]{\langle #1|}
\nc{\ket}[1]{|#1\rangle}
\nc{\braket}[1]{\langle #1 \rangle}
\nc{\Ep}{E_{\rm p}}
\nc{\tp}{t_{\rm p}}
\nc{\gtw}{g_{\rm tw}}
\nc{\kex}{\kappa_{\rm ex}}
\nc{\ki}{\kappa_{\rm i}}
\nc{\Dc}{\Delta_{\rm C}}
\nc{\Da}{\Delta_{\rm A}}

\begin{document}

\title{Microtoroidal cavity QED with fiber overcoupling and strong atom-field coupling: a single-atom quantum switch for coherent light fields}

\author{Scott Parkins}\email{s.parkins@auckland.ac.nz}
\affiliation{Department of Physics, 
             University of Auckland, Private Bag 92019, Auckland, New Zealand}
\author{Takao Aoki}\email{takao@waseda.jp}
\affiliation{Department of Applied Physics, 
             Waseda University, 3-4-1 Okubo, Shinjuku, Tokyo 169-8555, Japan}
\date{\today}

\begin{abstract}
We propose a scheme for single-atom, quantum control of the direction of propagation of a coherent field incident, via a tapered fiber, upon a microtoroidal whispering-gallery-mode (WGM) resonator. The scheme involves overcoupling of the fiber-taper to the resonator and strong coupling of an atom to the evanescent field of the WGM, i.e., an atom-field coupling that exceeds the total WGM linewidth. In contrast to previous, related schemes that operate in the bad-cavity regime, the proposed scheme can operate effectively with much stronger incident fields, while also preserving their coherent nature. It can also serve to prepare an entangled state of the atom and coherent optical pulses propagating in opposite directions along the fiber. We evaluate the fidelity of preparation of such a state taking into account absorption and atomic spontaneous emission and demonstrate that high fidelities should be possible with realistic parameters.
\end{abstract}


\maketitle

\section{Introduction}

Strong interactions between single atoms and single photons, engineered through the use of optical resonators or cavities (i.e., cavity quantum electrodynamics, or cavity QED), provide a means of realizing quantum dynamics and phenomena that are of both fundamental and practical interest \cite{Mabuchi02,Miller05,Parkins13}. On the fundamental side, for example, issues of quantum coherence, entanglement, and measurement can be addressed, while, on the practical side, capabilities for quantum state preparation and manipulation are of immediate relevance to quantum information processing. Indeed, in this latter context, cavity QED systems comprised of single atoms in optical Fabry-P\'erot cavities have already contributed an impressive range of experimental demonstrations, including deterministic single photon generation, state transfer between atoms and light, and conditional quantum dynamics \cite{Turchette95,Kuhn02,McKeever04,Keller04,Birnbaum05,Boozer07,Wilk07,Hijlkema07,Kubanek08,Barros09,Specht11,Ritter12,Reiserer13,Reiserer14}. These operations would be key building blocks for a quantum network in which atoms are used for storage of quantum information at local nodes of the network, photons (``flying qubits'') provide a means of distributing quantum information between distant nodes, and atom-photon interactions enable transfer between matter and light, and also manipulation, of the quantum information \cite{Kimble08}.

Despite this remarkable progress in experimental cavity QED, the ``conventional'' Fabry-P\'erot cavities used in the above-mentioned demonstrations do not lend themselves well to large scale networking. So, in the pursuit of scalable architectures for quantum networks based on effects in cavity QED and on propagating light fields, a variety of microfabricated (i.e., microchip-based) systems are now under active investigation; these include semiconductor quantum dots in micropillar, microdisk, or photonic crystal cavities (for recent reviews see, e.g., \cite{Reitzenstein12,Lodahl13}), cold atoms with monolithic WGM resonators \cite{Spillane05,Barclay06,Aoki06,Dayan08,Aoki09,Alton10,Junge13,OShea13,Shomroni14}, fiber-Fabry-P\'erot resonators \cite{Steinmetz06,Colombe07,Trupke07}, or photonic crystal waveguides \cite{Thompson13,Tiecke14,Yu14,Goban14}, and nanocrystals with microsphere cavities \cite{Thomas06,Park06}.

Key properties of ultra-low intrinsic cavity mode losses and high efficiency transfer of light fields into and out of the cavity modes are well established features of, in particular, microspheres, microdisks and microtoroids coupled via evanescent fields to tapered optical fibers \cite{Cai00,Armani03,Spillane03}. Additionally, the ultra-small volumes of the WGM's supported by these monolithic structures lead to very large single photon electric fields and, consequently, very large atom-field coupling strengths, $g$, which may exceed dissipative rates in the system. 
In fact, this regime of strong coupling cavity QED has now been demonstrated experimentally for both a single alkali atom in the evanescent field of a microtoroidal \cite{Aoki06} or microbottle \cite{Junge13} resonator and for a single quantum dot embedded in a microdisk resonator \cite{Srinivasan07a,Srinivasan08}.

An important aspect of these particular systems, related to the input-output coupling efficiency of photons, is the ability to ``tune'' the external coupling rate, $\kappa_{\rm ex}$, of the cavity modes to the optical fiber by adjusting the distance between the taper and the WGM microresonator. By setting $\kappa_{\rm ex}$ to be much larger than the rates for intrinsic cavity losses ($\ki$) and spontaneous emission of the atom or quantum dot ($\gamma$), one ensures that the fiber is indeed the dominant input-output channel in the system. 
Moreover, depending on the precise size of $\kappa_{\rm ex}$ relative to certain other coupling parameters, the microresonator cavity QED system can exhibit quite distinct regimes of operation with regards to effect on a light field propagating along an evanescently-coupled optical fiber. 

In particular, much interest has to date been focussed on operation under the condition of {\em critical coupling} between the fiber and the WGM microresonator \cite{Cai00,Aoki06,Junge13,OShea13}, i.e., $\kex=\kex^{\rm cr}$, whereby, in the absence of an atom, destructive interference prevents transmission of resonant light along the fiber, with, instead, light either reflected back in the opposite (backward) direction, dissipated through the intrinsic WGM losses, or transferred to a second evanescently-coupled fiber \cite{OShea13} (or, in general, some combination of these possibilities). A strongly coupled atom ($g>\{\kex,\ki,\gamma\}$) induces vacuum Rabi splitting of the cavity resonance and breaks the critical coupling condition, causing a finite transmission in the forward direction. However, such WGM microresonators support degenerate, counterpropagating modes, both of which couple to the atom. Identically-polarized, counterpropagating TE modes give rise to a pair of orthogonal standing-wave normal modes and the phase of one of these modes can always be chosen so that it has a node at the position of the atom. The presence of this uncoupled normal mode means that for TE fields the incident field can never be fully transmitted; in fact, the transmission is limited to 25\% of the incident field \cite{Aoki06b,Junge13} for the strong coupling regime $g>\{\kex,\ki,\gamma\}$. 

In the so-called ``bad cavity'' regime, for which $\kappa=\kex+\ki\gg g$ (but $g\gg\gamma$ so that $g^2/\kappa\gamma>1$), it is actually possible for this limit to be increased towards 100\% \cite{Dayan08}, but only for very weak incident fields, as the atom is typically saturated with extremely small photon numbers. Alternatively, it has been pointed out, and demonstrated experimentally, that for nontransversally-polarized WGM's (i.e., TM rather than TE fields) it is possible to realize a situation where the atom effectively couples to only a single, unidirectional WGM, such that strong vacuum Rabi splitting of this mode by the atom can in principle enable 100\% transmission along the fiber of light resonant with the bare WGM frequency \cite{Junge13,OShea13}.

In this paper, we describe a new and promising scheme which, under ideal conditions, also enables 100\% contrast in transmission along the fiber between the no-atom and single-atom cases, but which works despite the presence of an uncoupled, standing-wave normal mode of the microresonator (i.e., it works with transversally-polarized TE WGM's) and also enables control of the propagation of relatively strong light fields or, alternatively, of light pulses of significant photon number. The scheme requires the strong coupling regime of cavity QED, i.e., $g>\kappa,\gamma$, but, with regards to the fiber-microresonator coupling, operates in the {\em strongly over-coupled regime}, i.e., $\kex\gg\kex^{\rm cr}$. Under this condition, and in the absence of an atom, incident light is simply transmitted along the fiber. However, as we shall show, the effect of a single, strongly-coupled atom is now to effectively restore the condition of critical coupling and block the transmission of resonant light past the microresonator.

The present scheme can be viewed as complementing a previously studied, single-atom setup in the bad-cavity, over-coupled regime, demonstrated experimentally in \cite{Aoki09}, which exhibits the same contrast in transmission, but works only for weak incident light fields; in particular, in this regime the (two-level) atom can only reflect incident light efficiently provided that no more than one photon ever arrives within the effective atomic lifetime \cite{Carmichael93,Chang07,Rosenblum11}. This mechanism for reflection also yields strongly antibunched light in the reflected field, in contrast to the present scheme, which preserves coherent fields in reflection.

We begin in Section II with a description of our theoretical model, then in Section III present a linearized analysis, which allows a quite straightforward and transparent explanation of the present scheme. In Section IV we move beyond the linearized analysis to numerical solutions of the master equation for the system, enabling us to examine dependence on the atom-field coupling strength and on the coherent driving field strength. We connect the saturation behavior of the system with the semiclassical description of optical bistability of the cavity field, which allows us to quantify simply the regime of validity of the linear analysis. Given this knowledge, in Section V we consider the case of incident coherent-state pulses and give a conservative upper estimate for the mean photon number of a pulse that could in principle be controlled by the single-atom switch. In Section VI we then outline a scheme, involving an auxiliary (uncoupled) internal atomic level, which could allow the preparation of an entangled state of the atom and coherent state pulses propagating in opposite directions along the fiber. Using a linearized analysis, we evaluate the fidelity of preparation of such an entangled state for realistic parameters.

\section{Theoretical model}

\begin{figure}
  \includegraphics[scale=0.48]{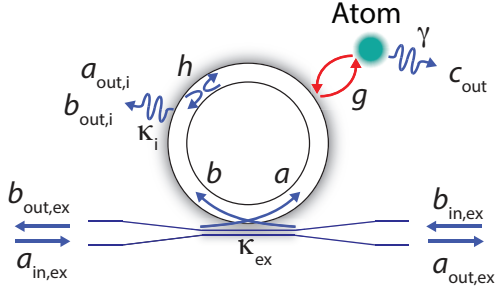}
  \caption{\label{fig:microtoroid} (Color online). Schematic of the microtoroid, atom, and tapered-fiber coupler. Whispering gallery modes, $a$ and $b$, couple with strength $g$ to an atom through their evanescent fields. The modes are coupled via intrinsic scattering at rate $h$ and suffer intrinsic loss at rate $\ki$ into the output fields $a_{\rm out,i}$ and $b_{\rm out,i}$. Evanescent coupling of the modes to the tapered fiber corresponds to an extrinsic loss at rate $\kex$, while atomic spontaneous emission into free space (output field $c_{\rm out}$) occurs at rate $\gamma$. Fields propagating along the fiber towards and away from the microtoroid are denoted by $\{ a_{\rm in,ex},b_{\rm in,ex}\}$ and $\{a_{\rm out,ex},b_{\rm out,ex}\}$, respectively. }
\end{figure}

A schematic of the system is shown in Fig.~\ref{fig:microtoroid}. The internal, counter-propagating 
cavity modes are described
in terms of the annihilation operators $a$ and $b$, while the external input and output fields are 
described by the operators $\{ a_{\rm in,ex},a_{\rm out,ex},b_{\rm in,ex},b_{\rm out,ex}\}$. The internal fields
suffer an intrinsic loss at the rate $\kappa_{\rm i}$ and an extrinsic loss 
at the rate $\kappa_{\rm ex}$ due to the fiber coupling.
An atom is assumed to couple to the evanescent fields of the intracavity (traveling-wave) modes
with a strength of magnitude $|g_{\rm tw}| = g_0^{\rm tw}\, f(r)$, where $r$ is the radial distance of the atom from the surface of the toroid and a characteristic form for the (real) function $f(r)$ is $f(r)={\rm e}^{-\alpha r}$ with $\alpha\sim 2\pi /\lambda$.

\subsection{Hamiltonian and master equation}

We consider a two-level atom with transition frequency $\omega_{\rm A}$ and described by the 
raising and lowering operators $\sigma^\pm$. The ``bare'' cavity mode (WGM)
frequency is $\omega_{\rm C}$, and the two counterpropagating modes are assumed to be coupled
(due, e.g., to scattering off imperfections) with a strength $h$. A coherent probe field of frequency
$\omega_{\rm p}$ in the input field $a_{\rm in,ex}$ drives the mode $a$ with strength $\Ep$. 
In a frame rotating at the probe frequency, the Hamiltonian for the system can be written in the form \cite{Aoki06b,Dayan08b,Domokos00,Rosenblit04,Srinivasan07b}
\begin{eqnarray}
H &=& \Delta_{\rm A}\sigma^+\sigma^- + \Delta_{\rm C} \left( a^\dag a + b^\dag b \right) 
+
h \left( a^\dag b + b^\dag a \right) \nonumber
\\
&& + \left( \Ep^\ast a + \Ep a^\dag \right) + \left( g_{\rm tw}^\ast a^\dag\sigma^- + g_{\rm tw} \sigma^+a \right)
\nonumber
\\
&& + \left( g_{\rm tw} b^\dag\sigma^- + g_{\rm tw}^\ast \sigma^+b \right)  ,
\end{eqnarray}
where, specifically, $g_{\rm tw}=g_0^{\rm tw}\, f(r) {\rm e}^{ikx}$, with $x$ the position of the atom around the circumference of the toroid, and
\begin{equation}
\Delta_{\rm A} = \omega_{\rm A} - \omega_{\rm p} \, , ~~~~
\Delta_{\rm C} = \omega_{\rm C} - \omega_{\rm p} .
\end{equation}
Introducing dissipation, the system can be described by the master equation
\begin{eqnarray}
\dot{\rho} &=& -{\rm i}\left[ H ,\rho \right] 
+ \kappa {\cal D}[a]\rho + \kappa {\cal D}[b]\rho +  \frac{\gamma}{2} {\cal D}[\sigma^-]\rho ,
\label{eq:ME}
\end{eqnarray}
where $\rho$ is the density operator for the atom-cavity system, $\kappa = \kappa_{\rm i} + \kappa_{\rm ex}$ is the field decay rate for the cavity modes, $\gamma$ is the atomic spontaneous emission
rate, and ${\cal D}[{\cal O}]\rho\equiv 2{\cal O}\rho{\cal O}^\dagger -  {\cal O}^\dagger{\cal O}\rho - \rho {\cal O}^\dagger{\cal O}$. For a fully quantum mechanical treatment of the system we can compute numerical solutions to the master equation (\ref{eq:ME}) using truncated number state bases for the cavity modes.

\subsection{Normal mode representation}

The Hamiltonian and master equation for the atom-cavity system can also be usefully expressed in terms of the normal modes of the cavity, which are defined by the operator combinations
\begin{equation}
A = \frac{a+b}{\sqrt{2}} \, , ~~~~~ B = \frac{a-b}{\sqrt{2}} \, .
\end{equation}
In particular, one can show that
\begin{eqnarray}
\dot{\rho} &=& -{\rm i}\left[ H^\prime ,\rho \right] 
+ \kappa {\cal D}[A]\rho + \kappa {\cal D}[B]\rho +  \frac{\gamma}{2} {\cal D}[\sigma^-]\rho ,
\label{eq:ME_AB}
\end{eqnarray}
where
\begin{eqnarray}
H^\prime &=& \Delta_{\rm A}\sigma^+\sigma^- + (\Delta_{\rm C}+h) A^\dag A 
+ (\Delta_{\rm C}-h) B^\dag B 
\nonumber
\\
&& + \frac{1}{\sqrt{2}} \left( \Ep^\ast A
+  \Ep A^\dag \right) + \frac{1}{\sqrt{2}} \left( \Ep^\ast B
+  \Ep B^\dag \right) \nonumber
\\
&&  + g_A \left( A^\dag\sigma^- + \sigma^+A \right) 
 - {\rm i}g_B \left( B^\dag\sigma^- - \sigma^+B \right) ,
\label{eq:H_normalmodes}
\end{eqnarray}
with
\begin{eqnarray}
g_A &=& \sqrt{2}\,{\rm Re}\{ g_{\rm tw}\} = \sqrt{2}\, g_0^{\rm tw}f(r)\cos(kx) ,
\\
g_B &=& \sqrt{2}\,{\rm Im}\{ g_{\rm tw}\} = \sqrt{2}\, g_0^{\rm tw} f(r)\sin(kx) .
\end{eqnarray}
So, depending on the position of the atom, coupling may occur predominantly
(or even exclusively) to only one of the two normal modes.

\subsection{Input and output fields}

From the theory of inputs and outputs in optical cavities \cite{CollettGardiner84_85,QuantumNoise}, the operators for the fields output from the two cavity modes into the fiber are given in terms of the input and intracavity field operators as
\begin{eqnarray}
a_{\rm out,ex}(t) &=& - a_{\rm in,ex}(t) + \sqrt{2\kappa_{\rm ex}}\, a(t) ,
\\
b_{\rm out,ex}(t) &=& - b_{\rm in,ex}(t) + \sqrt{2\kappa_{\rm ex}}\, b(t) ,
\end{eqnarray}
where $[a_{\rm in,ex}(t),a_{\rm in,ex}^\dagger (t^\prime)]=[a_{\rm out,ex}(t),a_{\rm out,ex}^\dagger (t^\prime)]=\delta (t-t^\prime )$ (and similarly for $b_{\rm in,ex}(t)$ and $b_{\rm out,ex}(t))$.
The particular form of the probe driving in the Hamiltonian corresponds to 
coherent amplitudes of the input fields given by
\begin{equation}
\langle a_{\rm in,ex} \rangle = - \frac{{\rm i}\Ep}{\sqrt{2\kappa_{\rm ex}}} , ~~~~~
\langle b_{\rm in,ex} \rangle = 0 ,
\end{equation}
and corresponding input photon fluxes of
\begin{equation}
\left|\langle a_{\rm in,ex}\rangle\right|^2 = 
\frac{\left|\Ep\right|^2}{2\kappa_{\rm ex}} \, , ~~~~~
\left|\langle b_{\rm in,ex}\rangle\right|^2 = 0 \, .
\end{equation}
The normalized forward and backward photon fluxes are therefore defined as
\begin{eqnarray}
T_{\rm F} = \frac{\langle a_{\rm out,ex}^\dag a_{\rm out,ex}\rangle_{\rm ss}}{\left|\Ep\right|^2/(2\kappa_{\rm ex})} \, , ~~~~
T_{\rm B} = \frac{\langle b_{\rm out,ex}^\dag b_{\rm out,ex}\rangle_{\rm ss}}{\left|\Ep\right|^2/(2\kappa_{\rm ex})} 
\end{eqnarray}
and the corresponding photon correlation functions as
\begin{eqnarray}
g_{\rm FF}^{(2)}(0) &=& \frac{\langle a_{\rm out,ex}^\dag a_{\rm out,ex}^\dag a_{\rm out,ex} a_{\rm out,ex}\rangle_{\rm ss}}{\left(\langle a_{\rm out,ex}^\dag a_{\rm out,ex}\rangle_{\rm ss}\right)^2} \, , 
\\
g_{\rm BB}^{(2)}(0) &=& \frac{\langle b_{\rm out,ex}^\dag b_{\rm out,ex}^\dag b_{\rm out,ex} b_{\rm out,ex}\rangle_{\rm ss}}{\left(\langle b_{\rm out,ex}^\dag b_{\rm out,ex}\rangle_{\rm ss}\right)^2} \, .
\end{eqnarray}

\section{Linearized analysis}

For sufficiently weak driving by a coherent probe field we can assume that the atom is
only very weakly excited and spends most of its time in the ground state, such that $\left[\sigma^-,\sigma^+\right]\rho \simeq \sigma^-\sigma^+\rho \simeq \rho$, or, effectively, $\left[\sigma^-,\sigma^+\right]\simeq 1$, and hence the atom behaves essentially like another harmonic oscillator. In this limit, all of the fields are coherent and we need only consider the mean field amplitudes, which obey the following equations of
motion:
\begin{widetext}
\begin{eqnarray}
\dot{\langle a\rangle} &=& -\left( \kappa + i\Delta_{\rm C} \right) \langle a\rangle - 
ih\langle b\rangle - i\Ep - ig_{\rm tw}^\ast \langle \sigma^-\rangle ,
\\
\dot{\langle b\rangle} &=& -\left( \kappa + i\Delta_{\rm C} \right) \langle b\rangle - 
ih\langle a\rangle - ig_{\rm tw} \langle \sigma^-\rangle ,
\\
\dot{\langle \sigma^-\rangle} &=& 
-\left( \gamma/2 + i\Delta_{\rm A} \right) \langle \sigma^-\rangle - 
ig_{\rm tw}\langle a\rangle - ig_{\rm tw}^\ast \langle b\rangle .
\end{eqnarray}
The steady state solutions for the cavity field amplitudes are
\begin{eqnarray}
\langle a\rangle_{\rm ss} &=& i\Ep\,
\frac{\left(\gamma/2+i\Delta_{\rm A}\right)\left[\left(\kappa+i\Delta_{\rm C}\right)\left(\gamma/2+i\Delta_{\rm A}\right)
+|g_{\rm tw}|^2\right]}
{\left[ ih\left(\gamma/2+i\Delta_{\rm A}\right)+{g_{\rm tw}^\ast}^2\right]
\left[ ih\left(\gamma/2+i\Delta_{\rm A}\right)+g_{\rm tw}^2\right]-
\left[\left(\kappa+i\Delta_{\rm C}\right)\left(\gamma/2+i\Delta_{\rm A}\right)+|g_{\rm tw}|^2\right]^2} \, ,
\label{eq:a_ss}
\\
\langle b\rangle_{\rm ss} &=& 
- \frac{ih\left(\gamma/2+i\Delta_{\rm A}\right)+g_{\rm tw}^2}
{\left(\kappa+i\Delta_{\rm C}\right)\left(\gamma/2+i\Delta_{\rm A}\right)+|g_{\rm tw}|^2}\, \langle a\rangle_{\rm ss} ,
\label{eq:b_ss}
\end{eqnarray}
\end{widetext}
and for the amplitudes of the cavity output fields we have
\begin{eqnarray}
\langle a_{\rm out,ex}\rangle_{\rm ss} &=& \frac{i\Ep}{\sqrt{2\kappa_{\rm ex}}} +
\sqrt{2\kappa_{\rm ex}}\,\langle a\rangle_{\rm ss} \, , 
\\
\langle b_{\rm out,ex}\rangle_{\rm ss} &=& \sqrt{2\kappa_{\rm ex}}\,\langle b\rangle_{\rm ss} \, .
\end{eqnarray}

\subsection{No atom}

In the absence of an atom ($g_{\rm tw}=0$) linear analysis is exact (all fields are coherent) and 
\begin{eqnarray}
\braket{a}_{\rm ss} &=& -\frac{i\Ep (\kappa +i\Dc)}{[\kappa+i(\Dc +h)][\kappa+i(\Dc -h)]} ,
\\
\braket{b}_{\rm ss} &=& - \frac{ih}{\kappa +i\Dc} \braket{a}_{\rm ss} .
\end{eqnarray}
For $\Delta_{\rm C}=\pm h$ one can then show that
\begin{eqnarray}
T_{\rm F} &=& \frac{(1-2\kappa_{\rm ex}/\kappa)^2+(4h^2/\kappa^2)(1-\kappa_{\rm ex}/\kappa)^2}{1+4h^2/\kappa^2} ,
\\
T_{\rm B} &=& \left( \frac{\kappa_{\rm ex}}{\kappa}\right)^2 \frac{4h^2/\kappa^2}{1+4h^2/\kappa^2} .
\end{eqnarray}
For $h=0$ (or $h\ll\kappa$) these reduce to (or are approximated by)
\begin{eqnarray}
T_{\rm F} &=& \left( 1 - 2\kappa_{\rm ex}/\kappa\right)^2 = \left( \frac{\kappa_{\rm i}-\kappa_{\rm ex}}{\kappa_{\rm i}+\kappa_{\rm ex}} \right)^2 ,
\\
T_{\rm B} &=& 0 .
\end{eqnarray}
Two cases are of particular interest to us. Firstly, critical coupling, for which $\kappa_{\rm ex}=\kex^{\rm cr}\equiv\sqrt{\kappa_{\rm i}^2+h^2}=\ki$ and hence $T_{\rm F}=0$. In this case, all of the incident light is dissipated through intrinsic loss in the microtoroid. Secondly, over-coupling, for which $\kappa_{\rm ex}\gg\{\kappa_{\rm i},h\}$, so $\kappa\simeq\kappa_{\rm ex}$ and $T_{\rm F}\simeq 1$; that is, virtually all of the light is simply transmitted along the fiber, past the microtoroid.

\subsection{Strongly-coupled atom}

\begin{figure}
  \centering
  \includegraphics[scale=0.45]{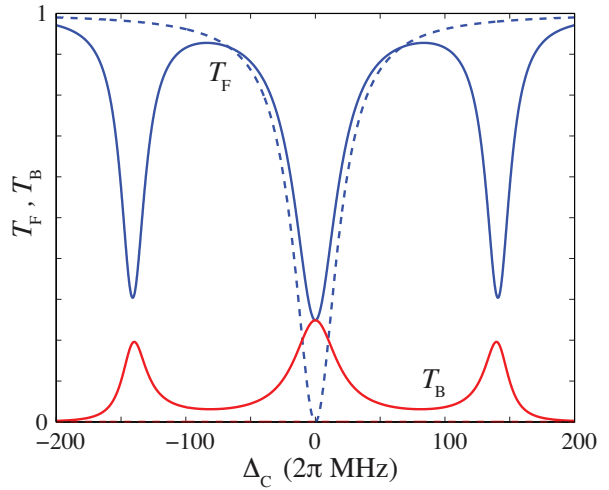}
  \caption{\label{fig:spec_cc} (Color online). Normalised power transmission $T_{\rm F}$ (blue) and reflection $T_{\rm B}$ (red) as a function of probe detuning in the weak-driving limit (using the linearised theory) for critical coupling ($\kex =\sqrt{\ki^2+h^2}$) and $\omega_{\rm A}=\omega_{\rm C}$. Parameters are $\{ \kex,\ki,h,\gamma\}/2\pi=\{10,10,0,5.2\}$~MHz. Solid lines are for $g_{\rm tw}/2\pi=100$~MHz ($\sin(kx)=0$) and dashed lines are for $g_{\rm tw}=0$. For large $\gtw$, the maximum increase in $T_{\rm F}(\Dc=0)$ is 0.25.}
\end{figure}

Now consider the case in which an atom couples strongly and resonantly (i.e., $\Da =\Dc$) to the microtoroid field modes, such that $|\gtw |\gg\{\kappa,\gamma,h,|\Dc |\}$. Then from (\ref{eq:a_ss}) and (\ref{eq:b_ss}) one finds
\begin{eqnarray}
\braket{a}_{\rm ss} &\simeq& -\frac{i\Ep/2}{\kappa+i[\Dc -h\cos(2kx)]} ,
\\
\braket{b}_{\rm ss} &\simeq& - {\rm e}^{2ikx} \braket{a}_{\rm ss} ,
\end{eqnarray}
which in turn give
\begin{eqnarray}
\braket{a_{\rm out,ex}}_{\rm ss} &\simeq& \frac{i\Ep}{\sqrt{2\kex}} \left\{ 1 - \frac{\kex}{\kappa+i[\Dc -h\cos(2kx)]} \right\},
\\
\braket{b_{\rm out,ex}}_{\rm ss} &\simeq& \frac{i\Ep}{\sqrt{2\kex}} {\rm e}^{2ikx} \frac{\kex}{\kappa+i[\Dc -h\cos(2kx)]} .
\end{eqnarray}
So, for $\Dc =h\cos(2kx)$ one has
\begin{eqnarray}
T_{\rm F} &\simeq & (1-\kex /\kappa)^2 = \left( \frac{\kappa_{\rm i}}{\kappa_{\rm i}+\kappa_{\rm ex}} \right)^2,
\label{eq:TFatom}
\\
T_{\rm B} &\simeq & (\kex /\kappa)^2 = \left( \frac{\kappa_{\rm ex}}{\kappa_{\rm i}+\kappa_{\rm ex}} \right)^2.
\label{eq:TBatom}
\end{eqnarray}
Now, for critical coupling, and with $h\ll\ki$ (so $\kex\simeq\ki$), we find that $T_{\rm F}\simeq T_{\rm B}\simeq 1/4$, and the increase (from zero) in forward and backward fluxes is limited to 25\% of the incident flux \cite{Aoki06b,Junge13}. This situation is illustrated in Fig.~\ref{fig:spec_cc}, where $T_{\rm F}$ and $T_{\rm B}$ are plotted as a function of probe detuning $\Dc$ under the conditions of critical coupling and with either $\gtw=0$ or $\gtw$ large. For the example shown ($\gtw$ real) the atom couples only to normal mode $A$ and the consequent normal mode splitting leads to a substantial reduction in $T_{\rm F}$ at $\Dc\simeq\pm\sqrt{2}\gtw$, i.e., at the vacuum Rabi sidebands. However, the transmission spectrum around $\Dc=0$ is dominated by the resonance of the uncoupled normal mode $B$, and $T_{\rm F}(\Dc=0)$ increases from zero by only 0.25 in the presence of the atom. Under more general conditions (i.e., $\gtw$ complex) the system, in the linear regime, amounts to three coupled oscillators, but the three resonances remain at $\Dc\simeq\pm\sqrt{2}\gtw$and $\Dc\simeq 0$ (for $h\simeq 0$), and  one still finds $T_{\rm F}(\Dc\simeq 0)\simeq 0.25$.

However, for strong over-coupling, $\kex\gg\ki$ (but, of course, with $\kex$ still much smaller than $|\gtw |$), it is apparent from (\ref{eq:TFatom}) and (\ref{eq:TBatom}) that $T_{\rm F}\simeq 0$ and $T_{\rm B}\simeq 1$ (cf. $T_{\rm F}\simeq 1$ and $T_{\rm B}\simeq 0$ with no atom). Hence, in this regime a single atom can act as a near-ideal switch for resonant or near-resonant light incident along the fiber. This is despite the presence of an intracavity normal mode that does not couple to the atom, which, as pointed out above, prevents this possibility in the case of critical coupling. In fact, if, for example (and without loss of generality), we consider the case in which $x=0$, so that the atom couples only to normal mode $A$, then the minimum in $T_{\rm F}$ actually occurs at $\Delta_{\rm C}=h$ (i.e., at $\omega_{\rm p}=\omega_{\rm C}-h$), which is the frequency of the {\em uncoupled} normal mode $B$.

\begin{figure}
 \centering  
  \includegraphics[scale=0.45]{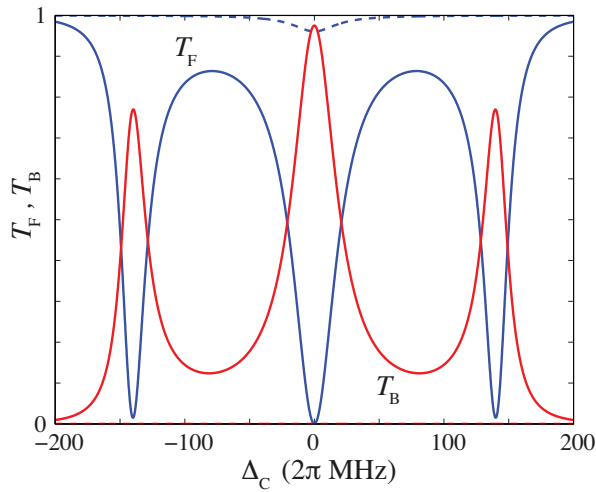}
  \caption{\label{fig:spec_oc} (Color online). Normalised power transmission $T_{\rm F}$ (blue) and reflection $T_{\rm B}$ (red) as a function of probe detuning in the weak-driving limit (using the linearised theory) for strong overcoupling ($\kex\gg\ki,h$) and $\omega_{\rm A}=\omega_{\rm C}$. Parameters are $\{ \kex,\ki,h,\gamma\}/2\pi=\{20,0.2,0,5.2\}$~MHz. Solid lines are for $g_{\rm tw}/2\pi=100$~MHz ($\sin(kx)=0$) and dashed lines are for $g_{\rm tw}=0$. For large $\gtw$, the maximum change in $T_{\rm F}(\Dc=0)$ is close to 1.}
\end{figure}

The overcoupled case is illustrated in Fig.~\ref{fig:spec_oc} and, to help understand the behavior on resonance ($\Dc=0$) for the case considered ($h=0$), we also give in Table~\ref{tabl:ampls} a summary of the approximate intracavity and output field amplitudes when $\sin (kx)=0$ and $\Delta_{\rm C}=h$. We observe that the presence of a strongly-coupled atom ($|\gtw |$ large) causes a redistribution of the intracavity field between the counterpropagating modes $a$ and $b$, in particular reducing by a factor of 2 the amplitude $\braket{a}_{\rm ss}$. This reduction now yields near-ideal destructive interference in the output field $a_{\rm out,ex}$ between the components $-a_{\rm in,ex}$ and $\sqrt{2\kex} a$, and consequently the incident light field is reflected along the fiber.

\begin{table}
  \begin{tabular}{c}
  \begin{tabular}{c c c c c c c}
    \multicolumn{7}{c}{Field amplitudes} \\
    \hline
    ~ & ~ & ~ &  $\gtw =0$ & ~ & ~ & $|\gtw |$ large \\
    \hline
    $\braket{a}_{\rm ss}$ & ~ & ~ & $-i\Ep /\kex$ & ~  & ~ & $-i\Ep /2\kex$  \\
    $\braket{b}_{\rm ss}$ & ~ & ~ & $0$ & ~  & ~ & $i\Ep /2\kex$  \\
    \hline
    $\braket{A}_{\rm ss}$ & ~ & ~ & $-i\Ep /\sqrt{2}\kex$ & ~  & ~ & $0$  \\
    $\braket{B}_{\rm ss}$ & ~ & ~ & $-i\Ep /\sqrt{2}\kex$ & ~  & ~ & $-i\Ep /\sqrt{2}\kex$  \\
    \hline
    $\braket{a_{\rm out,ex}}_{\rm ss}$ & ~ & ~ & $\braket{a_{\rm in,ex}}$ & ~  & ~ & $0$  \\
    $\braket{b_{\rm out,ex}}_{\rm ss}$ & ~ & ~ & $0$ & ~  & ~ & $-\braket{a_{\rm in,ex}}$  \\
    \hline
 \end{tabular}
   \end{tabular}
  \caption{Summary of approximate intracavity and output field amplitudes for operation in the overcoupled regime, $\kex\gg\{\ki ,h\}$, with $\gtw=0$ or $|\gtw |\gg\{\kappa,\gamma,h\}$. We take $\omega_{\rm A}=\omega_{\rm C}$, $\sin (kx)=0$,  and $\Dc=h$ (i.e., probe driving at the frequency of the uncoupled normal mode). Note that $\braket{a_{\rm in,ex}}=-i\Ep /\sqrt{2\kex}$.}
  \label{tabl:ampls}
\end{table}

\section{Numerical solutions of the master equation}

In this section we present results obtained from numerical solutions of the full quantum master equation (\ref{eq:ME}) in the regime of strong overcoupling ($\kex\gg\{\ki ,h\}$). Our main aim is to determine requirements on the system, in particular with regards to atom-field coupling strength and probe driving strength, for efficient operation as a switch for incident light fields.

\subsection{Variation with atom-field coupling strength}

\begin{figure}
  \centering
  \includegraphics[scale=0.42]{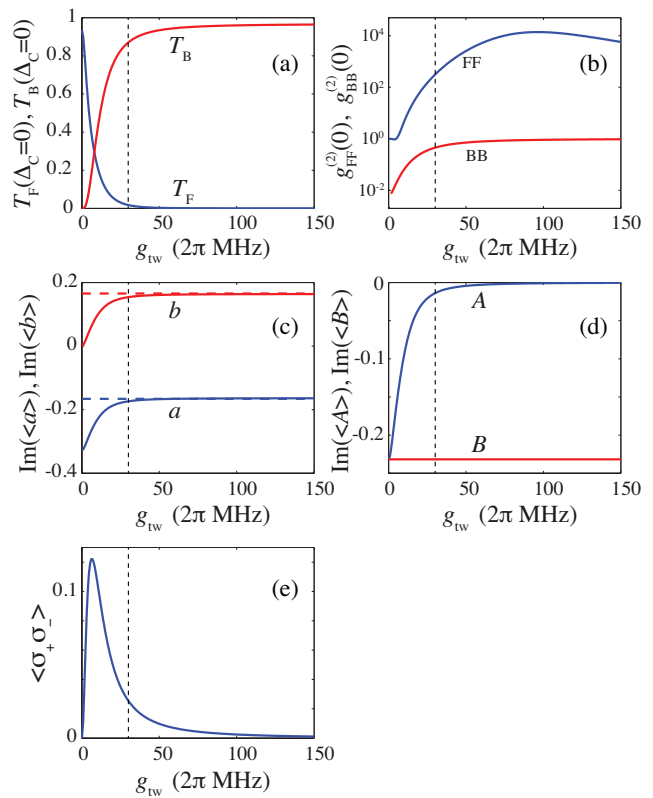}
  \caption{\label{fig:varying_gtw} (Color online). (a) Normalised power transmission $T_{\rm F}$ (blue) and reflection $T_{\rm B}$ (red), (b) output photon correlation functions $g_{\rm FF}^{(2)}(0)$ (blue) and $g_{\rm BB}^{(2)}(0)$ (red), (c) intracavity field amplitudes $\braket{a}$ (blue) and $\braket{b}$ (red), (d) normal mode amplitudes $\braket{A}$ (blue) and $\braket{B}$ (red), and (e) atomic excited state population as a function of atom-field coupling strength $\gtw$ for strong overcoupling ($\kex\gg\ki,h$) and $\omega_{\rm A}=\omega_{\rm C}=\omega_{\rm p}$ ($\Dc =\Da =0$). Parameters are $\{\kex,\ki,h,\gamma,E_{\rm p}\}/2\pi=\{30,0.5,0,5.2,10\}$~MHz. The horizontal dashed lines in (c) show the approximations to $\braket{a}$ and $\braket{b}$ from Table~\ref{tabl:ampls} for the limit of large $\gtw$. The vertical dashed line in each plot indicates the value of $\kex$.}
\end{figure}

First, we examine the behavior on resonance ($\Dc=\Da=0$) as a function of the atom-field coupling strength $\gtw$. Results for the power transmission and reflection in the fiber, the intracavity field amplitudes, and the atomic excited state population are presented in Fig.~\ref{fig:varying_gtw} for relatively weak driving field strength. Good operation of the atom-microtoroid system as a ``mirror'' for incident light is already observed once $\gtw >\kex$, with $T_{\rm F}(\Dc=0)\simeq 0$, $T_{\rm B}(\Dc=0)>0.9$ (Fig.~\ref{fig:varying_gtw}(a)), and $g_{\rm BB}^{(2)}(0)\simeq 1$ (Fig.~\ref{fig:varying_gtw}(b)), indicating that the reflected light remains essentially coherent. The atomic excited state population (Fig.~\ref{fig:varying_gtw}(e)) is seen to be very small for sufficiently large $\gtw$, hence keeping the effects of atomic spontaneous emission small, while the intracavity field amplitudes (Fig.~\ref{fig:varying_gtw}(c,d)) for $\gtw >\kex$ closely match the approximate results of Table \ref{tabl:ampls}.

We note from Fig.~\ref{fig:varying_gtw}(b) that the transmitted field $a_{\rm out,ex}$, although generally very weak, exhibits strong bunching over much of the range of $\gtw$. In terms of intracavity normal modes, this field is given by $a_{\rm out,ex}=-a_{\rm in,ex}+\sqrt{\kex}(A+B)$. For $\gtw$ real, the field component $\sqrt{\kex}B$ does not couple to the atom and is coherent, with an amplitude that, for $\kex\simeq\kappa$, essentially cancels with that of the coherent input field component, $-a_{\rm in,ex}$. The field component $\sqrt{\kex}A$, for $\gtw\gg\{\kex,\gamma\}$, typically has an amplitude even smaller than $-a_{\rm in,ex}+\sqrt{\kex}B$, but for resonant driving as considered here ($\Dc =\Da =0$) is strongly bunched; in particular, while absorption of a single photon by mode $A$ is very small in this regime, if one is absorbed then the probability of a second absorption is significantly enhanced, and, consequently, extreme bunching occurs in the photon statistics \cite{Kubanek08,Faraon08}, i.e., $\braket{A^\dag A^\dag AA}\gg\braket{A^\dag A}^2$.

\subsection{Variation with probe driving strength: saturation behavior}

\begin{figure}
  \centering
  \includegraphics[scale=0.42]{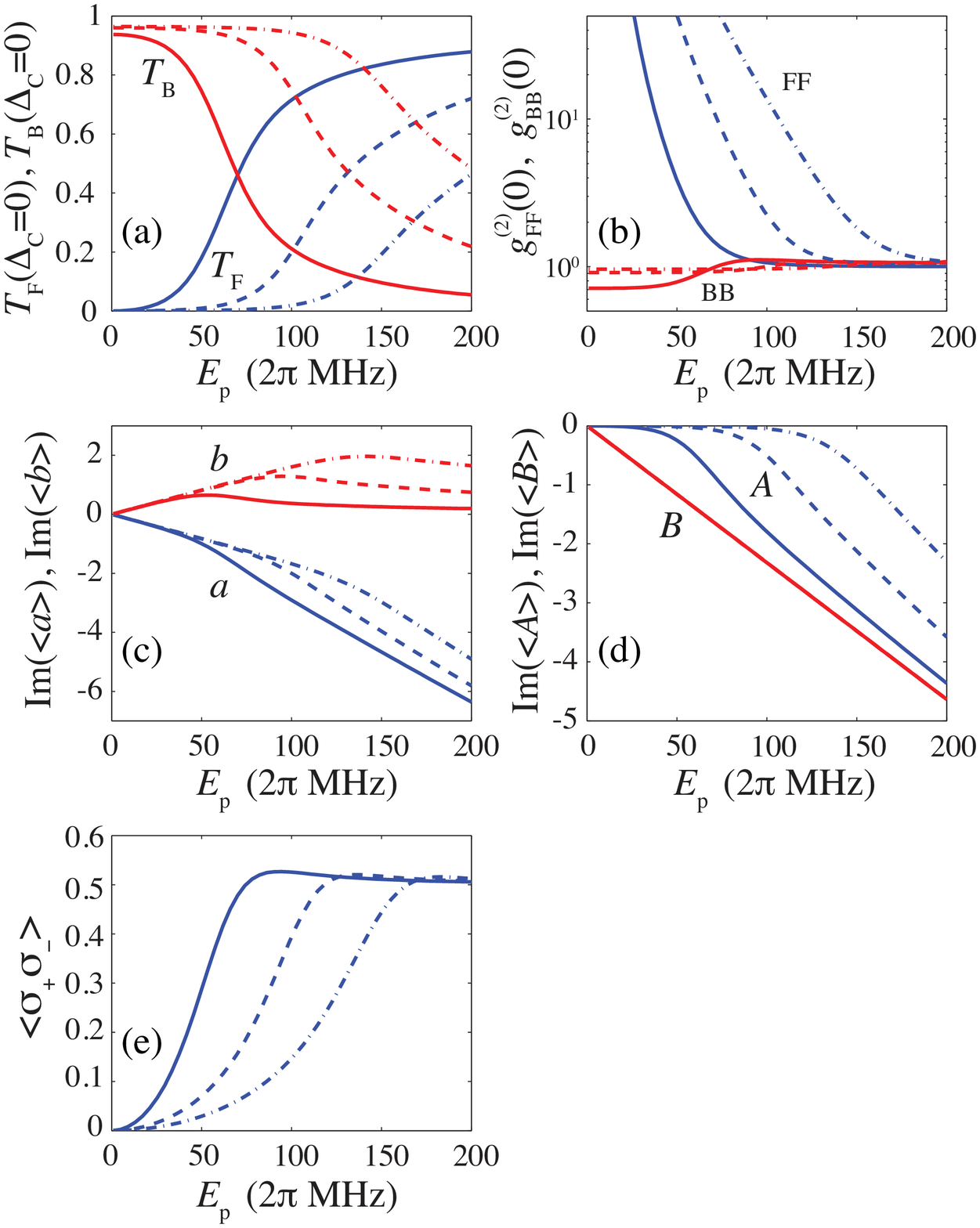}
  \caption{\label{fig:saturation} (Color online). (a) Normalised power transmission $T_{\rm F}$ (blue) and reflection $T_{\rm B}$ (red), (b) output photon correlation functions $g_{\rm FF}^{(2)}(0)$ (blue) and $g_{\rm BB}^{(2)}(0)$ (red), (c) intracavity field amplitudes $\braket{a}$ (blue) and $\braket{b}$ (red), (d) normal mode amplitudes $\braket{A}$ (blue) and $\braket{B}$ (red), and (e) atomic excited state population as a function of probe strength $E_{\rm p}$ for strong overcoupling ($\kex\gg\ki,h$) and $\omega_{\rm A}=\omega_{\rm C}=\omega_{\rm p}$ ($\Dc =\Da =0$), with $\gtw /2\pi=50$~MHz (solid), $100$~MHz (dashed), and $150$~MHz (dot-dashed). Other parameters are $\{\kex,\ki,h,\gamma\}/2\pi=\{30,0.5,0,5.2\}$~MHz.}
\end{figure}

For sufficiently strong driving field strength the atom begins to saturate and the desired behavior just discussed starts to break down. This is illustrated in Fig.~\ref{fig:saturation}, where the key quantities of interest are plotted now as a function of probe strength $E_{\rm p}$ for several values of the atom-field coupling strength $\gtw$. With the onset of saturation, $T_{\rm F}(\Dc=0)$ ($T_{\rm B}(\Dc=0)$) (Fig.~\ref{fig:saturation}(a)) increases (decreases) steadily from zero (one). The probe strength at which this onset occurs increases approximately linearly with the value of $\gtw$; we investigate this further below. 
As an aside, we note the interesting result that a small steady-state atomic inversion ($\braket{\sigma_+\sigma_-}>0.5$ or $\braket{\sigma_z}>0$) occurs over a range of $E_{\rm p}$ depending on the magnitude of $\gtw$ (Fig.~\ref{fig:saturation}(e)). 

\begin{figure}
  \centering
  \includegraphics[scale=0.42]{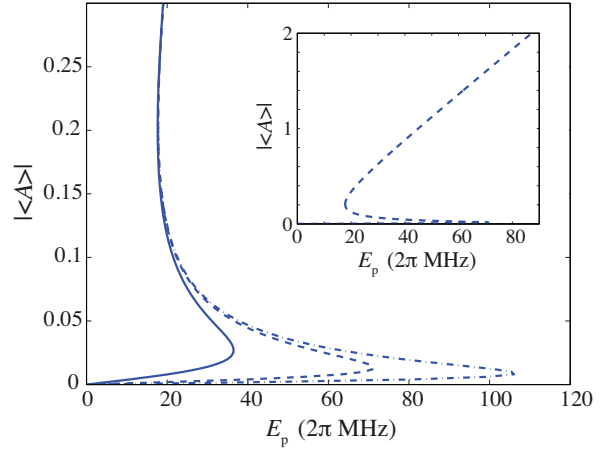}
  \caption{\label{fig:bistability} (Color online). Optical bistability curves for $\gtw /2\pi=50$~MHz (solid), $100$~MHz (dashed), and $150$~MHz (dot-dashed) with strong overcoupling ($\kex\gg\ki,h$) and $\omega_{\rm A}=\omega_{\rm C}=\omega_{\rm p}$ ($\Dc =\Da =0$). The inset shows the case $\gtw /2\pi=100$~MHz on a larger vertical scale. Other parameters are $\{\kex,\ki,h,\gamma\}/2\pi=\{30,0.5,0,5.2\}$~MHz.}
\end{figure}

If we consider the case $\sin (kx)=0$ (as in Fig.~\ref{fig:saturation}), then the atom couples only to the normal mode $A=(a+b)/\sqrt{2}$. If we further assume that $h=0$ and $\Dc=\Da=0$, then one can derive in a semiclassical approximation (whereby, for example, we set $\braket{\sigma_+a}=\braket{\sigma_+}\braket{a}$ in the equations of motion for $\braket{\sigma_+}$, $\braket{\sigma_-}$, $\braket{a^\dagger}$, $\braket{a}$, and $\braket{\sigma_z}$) the following equation relating the driving field strength, $\Ep$, to the normal mode amplitude, $\braket{A}$:
\begin{align}
|Y| = |X| \left( 1 + \frac{4C}{1+2|X|^2} \right) .
\end{align}
Here,
\begin{align}
Y = \frac{i\Ep}{\kappa\sqrt{2n}}  , ~~~ X = \frac{\braket{A}}{\sqrt{n}}  , ~~~ n=\frac{\gamma^2}{8g_{\rm tw}^2} , ~~~ C=\frac{g_{\rm tw}^2}{\kappa\gamma} .
\end{align}
This is simply the equation of optical bistability for the normal mode $A$. Bistability curves, in terms of the normal mode amplitude, $|\braket{A}|$, and the driving field amplitude, $\Ep$, are shown in Fig.~\ref{fig:bistability}.

By setting $d|Y|/d|X|=0$ one can find the turning points of these curves. For $C\gg 1$ (as is the case for the situation we consider), these turning points are located at $|X|\simeq\sqrt{2C}$ and $|X|\simeq\sqrt{1/2}$, with corresponding driving field strengths 
\begin{align}
\Ep = \sqrt{2\kappa\gamma} ~~~ {\rm and} ~~~ \Ep=g_{\rm tw} \sqrt{\frac{1}{8}}\, \frac{1+2C}{C} \simeq \frac{g_{\rm tw}}{\sqrt{2}} \, .
\end{align}
The value of $\Ep$ at the lower turning point, $\Ep\simeq g_{\rm tw}/\sqrt{2}$, agrees well with the value, observed in the numerical simulations of the full quantum model, at which the onset of saturation occurs, i.e., at which $\braket{A}$ (Fig.~\ref{fig:saturation}(d)) and $T_{\rm F}(\Dc=0)$ start to deviate significantly from zero.

\begin{figure}
  \centering
  \includegraphics[scale=0.46]{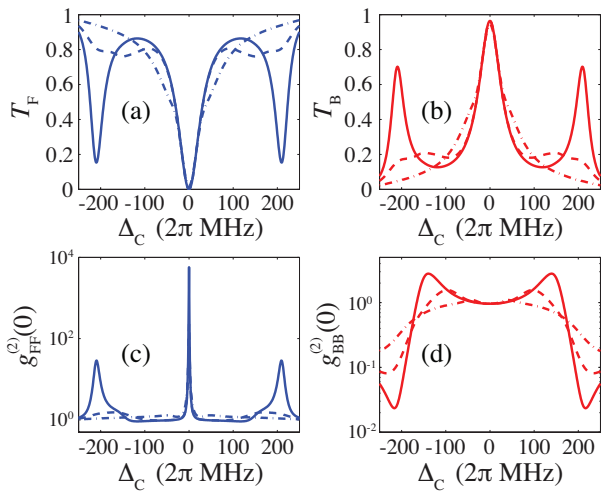}
  \caption{\label{fig:spec_Ep} (Color online). Normalised power (a) transmission $T_{\rm F}$ and (b) reflection $T_{\rm B}$, and output photon correlation functions (c) $g_{\rm FF}^{(2)}(0)$ and (d) $g_{\rm BB}^{(2)}(0)$, for $\gtw /2\pi=150$~MHz with strong overcoupling ($\kex\gg\ki,h$), $\omega_{\rm A}=\omega_{\rm C}$, and $\Ep /2\pi =10$~MHz (solid), $50$~MHz (dashed), and $100$~MHz (dot-dashed). Other parameters are $\{\kex,\ki,h,\gamma\}/2\pi=\{30,0.5,0,5.2\}$~MHz.}
\end{figure}

Finally, to further investigate the effects of saturation, in Fig.~\ref{fig:spec_Ep} we plot the transmission and reflection as a function of probe detuning (Figs.~\ref{fig:spec_Ep}(a,b)), as well as the output photon correlation functions (Figs.~\ref{fig:spec_Ep}(c,d)), for several values of the probe strength, with $\gtw/2\pi = 150$~MHz. While properties associated with the vacuum Rabi sidebands around $\Dc\simeq\sqrt{2}\gtw$ are drastically affected by saturation, the resonance feature centered at $\Dc=0$ (and central to the operation of the system as a switch for resonant light) is very robust to increasing probe strength.

\section{Coherent-state pulse reflection}

As shown above, for strong atom-field coupling (i.e., for $C\gg 1$), we find a maximum driving field strength of $\Ep\simeq g_{\rm tw}/\sqrt{2}$ before saturation effects start to occur. The incident photon flux,
\begin{align}
{\cal F} = \frac{\Ep^2}{2\kappa_{\rm ex}} ,
\end{align}
is therefore restricted to a maximum value
\begin{align}
{\cal F}_{\rm sat} \simeq \frac{g_{\rm tw}^2}{4\kappa_{\rm ex}} ,
\end{align}
if saturation effects are to be avoided. If we take, for example, $g_{\rm tw}=2\pi\cdot 100$~MHz and $\kappa_{\rm ex}=2\pi\cdot 30$~MHz, then ${\cal F}_{\rm sat}\simeq 520~(\mu {\rm s})^{-1}$, corresponding to a maximum input power, at 852~nm, of ${\cal P}_{\rm sat}\simeq 0.12$~nW.

Now consider an incident Gaussian pulse with a maximum photon flux of ${\cal F}_{\rm max}$, i.e., an incident flux profile of the form
\begin{align}
{\cal F}(t) = {\cal F}_{\rm max} \exp \left( -\frac{t^2}{2t_{\rm p}^2} \right) ,
\end{align}
where $2.35t_{\rm p}$ gives the FWHM of the pulse. If we consider the transmission and reflection spectra of Fig.~\ref{fig:spec_Ep}, then the width of the central resonance is approximately $2\kappa\simeq 2\kappa_{\rm ex}$. The frequency bandwidth of the incident pulse should be much less than this width; that is, we require
\begin{align}
\frac{2.35}{t_{\rm p}} \ll 2\kappa_{\rm ex} ,
\end{align}
which, for $\kappa_{\rm ex}=2\pi\cdot 30$~MHz, gives $t_{\rm p}\gg 6.2~{\rm ns}$. Choosing $t_{\rm p}=310~{\rm ns}$ and ${\cal F}_{\rm max}={\cal F}_{\rm sat}/2\simeq 260~(\mu {\rm s})^{-1}$, the (average) total photon number in the pulse is
\begin{align}
\bar{N}_{\rm p} = \int_{-\infty}^\infty {\cal F}(t)\, dt = {\cal F}_{\rm max} \sqrt{2\pi}\, t_{\rm p} \simeq 200 .
\end{align}
This suggests that our single-atom system could act effectively as a ``switch'' for an incident light pulse of significant photon number.
This in turn raises interesting possibilities for the generation, via the intrinsically quantum nature of the single-atom switch, of entangled states of an atom and propagating, coherent-state light fields, which we consider in the following section.

Before doing so, our scheme for pulse reflection should be contrasted with schemes based upon a single two-level emitter or atom coupled strongly to light fields propagating in a waveguide (e.g., nanowire or fiber) \cite{Chang07,Aoki09}. In these cases, the effective output field in the forward direction takes the form $a_{\rm out,ex}=-a_{\rm in,ex}+\sqrt{\Gamma}\sigma_-$, where $\Gamma$ is the Purcell-effect-enhanced spontaneous emission rate of the emitter. Given the two-level nature of the emitter, efficient reflection of light can only occur ``one photon at a time,'' and saturation of the emitter, marking the onset of an increase in transmission, corresponds to a single photon within a pulse of duration $\Gamma^{-1}$
(or, alternatively, within the effective bandwidth of the device) \cite{Carmichael93,Chang07}. 
In the present scheme, however, saturation corresponds to a photon number $\sqrt{2\pi} {\cal F}_{\rm sat}/\kappa_{\rm ex}\sim (\gtw /\kappa_{\rm ex})^2$ within a (Gaussian) pulse of duration $\kappa_{\rm ex}^{-1}$ (the inverse of the effective bandwidth of our device), which may clearly be much larger than one and is not limited by the two-level nature of the atom. 
That this is possible is evident from consideration of the output field for our scheme, written in the form $a_{\rm out,ex}=-a_{\rm in,ex}+\sqrt{\kex}(A+B)$. An atom coupled strongly to normal mode $A$ takes this mode far out of resonance, reducing its amplitude essentially to zero and the output field to $a_{\rm out,ex}\simeq -a_{\rm in,ex}+\sqrt{\kex}B$, where $B$ is the orthogonal, uncoupled normal mode, which for coherent input fields is itself simply a coherent field. The amplitude of $B$ is not constrained and is such as to give near ideal destructive interference in the output field $a_{\rm out,ex}$ (assuming, of course, operation within the saturation restrictions of the system).

\section{Entangled-path coherent-state preparation}

Let $\ket{\rm g}$ and $\ket{\rm e}$ denote the atomic states coupled via the microtoroid field modes, and let $\ket{\rm g'}$ denote an auxiliary internal atomic state (for example, in another hyperfine level) that does not couple to the field modes. If the atom is prepared in a superposition of $\ket{\rm g}$ and $\ket{\rm g'}$, and a coherent state pulse, denoted by $\ket{\alpha}_{a_{\rm in,ex}}$, is incident along the fiber (in input field $a_{\rm in,ex}$), then under ideal operation of the quantum switch, the following transformation would be implemented,
\begin{align}
&\ket{\alpha}_{a_{\rm in,ex}}  \ket{0}_{b_{\rm in,ex}} \frac{1}{\sqrt{2}}\left( \ket{\rm g}+\ket{\rm g'} \right) \nonumber
\\
& ~~~\longrightarrow \ket{\Phi} = \frac{1}{\sqrt{2}} \ket{0}_{a_{\rm out,ex}} \ket{-\alpha}_{b_{\rm out,ex}} \ket{\rm g}  \nonumber
\\
& ~~~~~~~~~~~~~~~~~~~+\frac{1}{\sqrt{2}} \ket{\alpha}_{a_{\rm out,ex}} \ket{0}_{b_{\rm out,ex}} \ket{\rm g'}  ,
\end{align}
where $\ket{0}_{b_{\rm in}}$ denotes a vacuum state in $b_{\rm in}$. That is, the atomic state becomes entangled with a coherent state pulse (of potentially quite large photon number, as shown in the previous section) propagating in either the forward or backward direction along the fiber. With a suitable (unitary) rotation and projective measurement of the atomic state, this system could therefore be used to prepare entangled coherent states, which are of significant interest in the contexts of, e.g., quantum information processing and quantum metrology 
\cite{Ourjoumtsev09,Sanders12,Zhang13,Knott14}. However, such states are by nature extremely sensitive (i.e., fragile) to uncontrolled losses, which in the present system arise from intrinsic losses (i.e., absorption) in the resonator modes and from atomic spontaneous emission into free space. More detailed investigation of the effect of these losses, as well as the influence of key parameter values, on entangled-state preparation is therefore important and now follows. 

\subsection{Input pulse}

We consider a coherent-state, Gaussian optical pulse incident along the fiber, which we describe by \cite{Wang05}
\begin{widetext}
\begin{align}
\ket{\alpha}_{a_{\rm in,ex}} = \exp \left( -|\alpha|^2/2\right) \exp\left[ \int_{-\infty}^\infty \braket{a_{\rm in,ex}(t)} a_{\rm in,ex}^\dagger (t) dt\right] \ket{0}_{a_{\rm in,ex}} ,
\end{align}
\end{widetext}
where
\begin{align}
\braket{a_{\rm in,ex}(t)} &= - \frac{i\Ep}{\sqrt{2\kex}} \exp \left( - \frac{t^2}{4\tp^2} \right),
\end{align}
or, equivalently,
\begin{align}
\braket{\tilde{a}_{\rm in,ex}(\omega)} &= \frac{1}{\sqrt{2\pi}} \int_{-\infty}^\infty {\rm e}^{-i\omega t} \braket{a_{\rm in,ex}(t)} dt \nonumber
\\
&= - \frac{i\Ep\tp}{\sqrt{\kex}} \exp \left( -\omega^2\tp^2 \right) ,
\end{align}
and
\begin{align}
\int_{-\infty}^\infty |\braket{a_{\rm in,ex}(t)}|^2 dt &= \int_{-\infty}^\infty |\braket{\tilde{a}_{\rm in,ex}(\omega)}|^2 d\omega \nonumber
\\
&= \bar{N}_{\rm p} \equiv |\alpha|^2 .
\end{align}

\subsection{Linear (coherent state) analysis}

If we assume operation of the system in the linear regime, i.e., that the driving strength is sufficiently small ($\Ep <\gtw/\sqrt{2}$), then we can solve operator equations of motion for the resonator and atomic modes exactly and derive expressions for all of the output field operators (in frequency space) in terms of $\tilde{a}_{\rm in,ex}(\omega)$ (Appendix A). Furthermore, in a linear system all fields remain coherent, so, for the atom in state $\ket{\rm g}$, we can write the output state of our system as a product of coherent states in each of the possible output channels \cite{CohStateApprox}, i.e., 
\begin{align}
\ket{\psi} = \ket{\alpha_{\rm ex}}_{a_{\rm out,ex}} \ket{\alpha_{\rm i}}_{a_{\rm out,i}} \ket{\beta_{\rm ex}}_{b_{\rm out,ex}} \ket{\beta_{\rm i}}_{b_{\rm out,i}} \ket{\eta}_{c_{\rm out}} ,
\end{align}
with, for example (see Appendix A),
\begin{widetext}
\begin{align}
\ket{\alpha_{\rm ex}}_{a_{\rm out,ex}} = \exp \left( -|\alpha_{\rm ex}|^2/2\right) \exp\left[ \int_{-\infty}^\infty \braket{a_{\rm out,ex}(t)} a_{\rm out,ex}^\dagger (t) dt\right] \ket{0}_{a_{\rm out,ex}} ,
\end{align}
\end{widetext}
where
\begin{align}
\braket{a_{\rm out,ex}(t)} &= \frac{1}{\sqrt{2\pi}} \int_{-\infty}^\infty {\rm e}^{i\omega t} \braket{\tilde{a}_{\rm out,ex}(\omega)} d\omega 
\\
&= \frac{1}{\sqrt{2\pi}} \int_{-\infty}^\infty {\rm e}^{i\omega t} t_{\rm ex}(\omega)\braket{\tilde{a}_{\rm in,ex}(\omega)} d\omega ,
\end{align}
\begin{align}
|\alpha_{\rm ex}|^2 = \int_{-\infty}^\infty |t_{\rm ex}(\omega)|^2 |\braket{\tilde{a}_{\rm in,ex}(\omega)}|^2 d\omega ,
\end{align}
and $t_{\rm ex}(\omega)$ is also given in Appendix A. 
Note that $c_{\rm out}$ denotes the output field from the atom into free space (spontaneous emission).
Similarly, for the atom in state $\ket{\rm g'}$ we can write
\begin{align}
\ket{\psi^\prime} = \ket{\alpha_{\rm ex}^{(0)}}_{a_{\rm out,ex}} \ket{\alpha_{\rm i}^{(0)}}_{a_{\rm out,i}} \ket{\beta_{\rm ex}^{(0)}}_{b_{\rm out,ex}} \ket{\beta_{\rm i}^{(0)}}_{b_{\rm out,i}} \ket{\eta^{(0)}}_{c_{\rm out}} ,
\end{align}
where now, for example,
\begin{align}
|\alpha_{\rm ex}^{(0)}|^2 = \int_{-\infty}^\infty |t_{\rm ex}^{(0)}(\omega)|^2 |\braket{\tilde{a}_{\rm in,ex}(\omega)}|^2 d\omega ,
\end{align}
with
\begin{align}
t_{\rm ex}^{(0)}(\omega) = \left. t_{\rm ex}(\omega)\right|_{\gtw=0} .
\end{align}
Hence, for the atom in an equal superposition of $\ket{\rm g}$ and $\ket{\rm g'}$, the output state of the system is given by
\begin{align}
\ket{\Psi} = \frac{1}{\sqrt{2}} \left( \ket{\rm g}\ket{\psi} + \ket{\rm g^\prime}\ket{\psi^\prime} \right) .
\end{align}

\subsection{Reduced density operator}

Tracing over the output fields $a_{\rm out,i}$, $b_{\rm out,i}$, and $c_{\rm out}$, we obtain a reduced density operator for the fiber output fields (and atom) only, which takes the form
\begin{align}
\rho &=\frac{1}{2} \ket{\rm g} \ket{\alpha_{\rm ex}} \ket{\beta_{\rm ex}} \bra{\beta_{\rm ex}} \bra{\alpha_{\rm ex}} \bra{\rm g} \nonumber
\\
&+\,\frac{1}{2} \ket{\rm g^\prime} \ket{\alpha_{\rm ex}^{(0)}} \ket{\beta_{\rm ex}^{(0)}} \bra{\beta_{\rm ex}^{(0)}} \bra{\alpha_{\rm ex}^{(0)}} \bra{\rm g^\prime} \nonumber
\\
&+\, \frac{\xi}{2} \ket{\rm g} \ket{\alpha_{\rm ex}} \ket{\beta_{\rm ex}} \bra{\beta_{\rm ex}^{(0)}} \bra{\alpha_{\rm ex}^{(0)}} \bra{\rm g^\prime} \nonumber
\\
&+\, \frac{\xi^\ast}{2} \ket{\rm g^\prime} \ket{\alpha_{\rm ex}^{(0)}} \ket{\beta_{\rm ex}^{(0)}} \bra{\beta_{\rm ex}} \bra{\alpha_{\rm ex}} \bra{\rm g} ,
\end{align}
where, for brevity, we have dropped the output field subscripts; states associated with output channels $a_{\rm out,ex}$ and $b_{\rm out,ex}$ can be distinguished by the use of $\{\alpha_{\rm ex},\alpha_{\rm ex}^{(0)}\}$ or $\{\beta_{\rm ex},\beta_{\rm ex}^{(0)}\}$, respectively. The parameter $\xi$ appearing in the reduced density operator is given by 
\begin{align}
\xi =  \braket{\alpha_{\rm i}^{(0)}|\alpha_{\rm i}}_{a_{\rm out,i}} \braket{\beta_{\rm i}^{(0)}|\beta_{\rm i}}_{b_{\rm out,i}} \braket{\eta^{(0)}|\eta}_{c_{\rm out}} ,
\end{align}
where one can show (noting that, in fact, $\beta_{\rm i}^{(0)} =0$ for $h=0$, and $\eta^{(0)}=0$) that
\begin{widetext}
\begin{align}
\braket{\alpha_{\rm i}^{(0)}|\alpha_{\rm i}}_{a_{\rm out,i}} &= \exp \left[ -\frac{1}{2} \int_{-\infty}^\infty  \left( |t_{\rm i}(\omega)|^2 + |t_{\rm i}^{(0)}(\omega)|^2 - 2t_{\rm i}^{(0)}(\omega)^\ast t_{\rm i}(\omega) \right) |\braket{\tilde{a}_{\rm in,ex}(\omega)}|^2 d\omega \right] ,
\\
\braket{\beta_{\rm i}^{(0)}|\beta_{\rm i}}_{b_{\rm out,i}} &=\exp \left[ -\frac{1}{2} \int_{-\infty}^\infty |r_{\rm i}(\omega)|^2 |\braket{\tilde{a}_{\rm in,ex}(\omega)}|^2 d\omega \right] \equiv \exp \left[ -\frac{1}{2} |\beta_{\rm i}|^2 \right] ,
\\
\braket{\eta^{(0)}|\eta}_{c_{\rm out}} &= \exp \left[ -\frac{1}{2} \int_{-\infty}^\infty |s(\omega)|^2 |\braket{\tilde{a}_{\rm in,ex}(\omega)}|^2 d\omega \right] \equiv \exp \left[ -\frac{1}{2} |\eta|^2 \right] ,
\end{align}
\end{widetext}
with $t_{\rm i}(\omega)$, $r_{\rm i}(\omega)$, and $s(\omega)$ given in Appendix A. Note that these are overlap factors between the states of these output channels with the atom either in state $\ket{\rm g}$ or state $\ket{\rm g'}$. The smaller these factors are, the more ``distinguishable'' are the components of the desired entangled state from information carried by these channels, and the worse the expected fidelity of preparation of the entangled coherent state (i.e., the reduced density operator approaches a mixed state in the limit $\xi\rightarrow 0$).

\subsection{Fidelity of entangled state preparation}

Finally, the fidelity of preparation of the ideal output state $\ket{\Phi}$ is given by
\begin{align}
F &= \bra{\Phi} \rho \ket{\Phi} \nonumber
\\
&= \frac{1}{4} |\braket{0|\alpha_{\rm ex}}|^2 |\braket{-\alpha |\beta_{\rm ex}}|^2  + \frac{1}{4} |\braket{\alpha |\alpha_{\rm ex}^{(0)}}|^2 |\braket{0 |\beta_{\rm ex}^{(0)}}|^2 \nonumber
\\
& +\, \frac{\xi}{4} \braket{0|\alpha_{\rm ex}} \braket{-\alpha |\beta_{\rm ex}} \braket{\alpha_{\rm ex}^{(0)} |\alpha} \braket{\beta_{\rm ex}^{(0)}|0} \nonumber
\\
& +\, \frac{\xi^\ast}{4} \braket{\alpha_{\rm ex}|0} \braket{\beta_{\rm ex}|-\alpha} \braket{\alpha |\alpha_{\rm ex}^{(0)}} \braket{0|\beta_{\rm ex}^{(0)}} ,
\label{eq:F}
\end{align}
where
\begin{align}
\braket{0|\alpha_{\rm ex}} &= \exp\left( -\frac{1}{2} |\alpha_{\rm ex}|^2 \right) ,
\\
\braket{0 |\beta_{\rm ex}^{(0)}} &= \braket{0|0} =1 ,
\end{align}
and
\begin{widetext}
\begin{align}
\braket{-\alpha |\beta_{\rm ex}} &= \exp \left[ -\frac{1}{2} \int_{-\infty}^\infty \left( 1 + |r_{\rm ex}(\omega)|^2 + 2r_{\rm ex}(\omega) \right) |\braket{\tilde{a}_{\rm in,ex}(\omega)}|^2 d\omega \right] ,
\\
\braket{\alpha |\alpha_{\rm ex}^{(0)}} &= \exp \left[ -\frac{1}{2} \int_{-\infty}^\infty \left( 1 + |t_{\rm ex}^{(0)}(\omega)|^2 - 2t_{\rm ex}^{(0)}(\omega) \right) |\braket{\tilde{a}_{\rm in,ex}(\omega)}|^2 d\omega \right] ,
\end{align}
\end{widetext}
with $r_{\rm ex}(\omega)$ given in Appendix A.

\begin{figure}[t]
  \centering
  \includegraphics[scale=0.46]{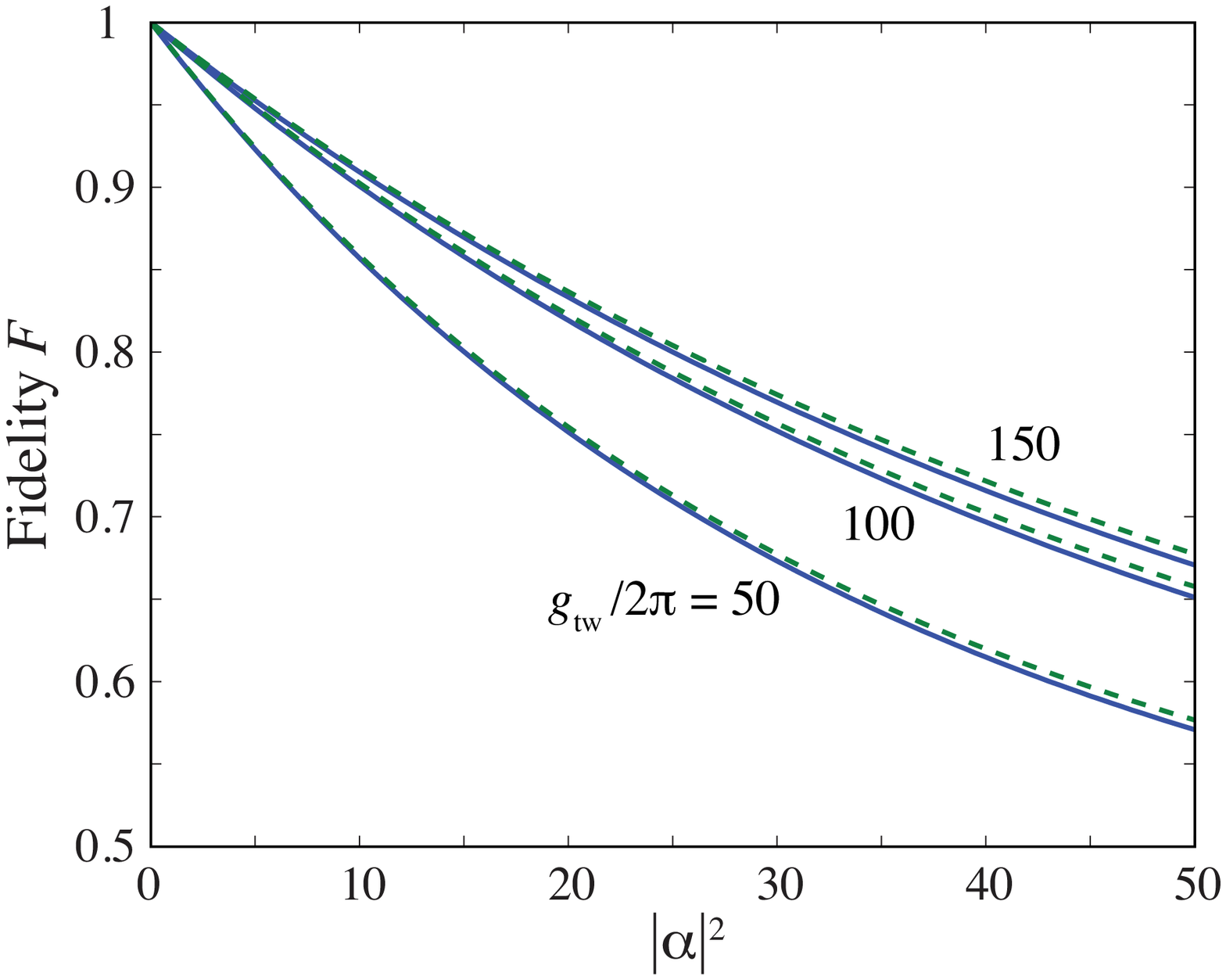}
  \caption{\label{fig:F_g} (Color online). Fidelity, $F$, of entangled-path coherent state preparation as a function of the mean photon number, $|\alpha|^2$, in the incident pulse for $\gtw /2\pi =\{50,100,150\}$~MHz (curves are labelled by these values in units of MHz). Dashed lines are the approximate expression (\ref{eq:F_approx}). Other parameters are $\{\kex,\ki,h,\gamma\}/2\pi=\{30,0.5,0,5.2\}$~MHz, $t_{\rm p}=318$~ns ($\kappa t_{\rm p}\simeq 60$), and $\omega_{\rm A}=\omega_{\rm C}$.}
\end{figure}

\begin{figure}[t]
  \centering
  \includegraphics[scale=0.46]{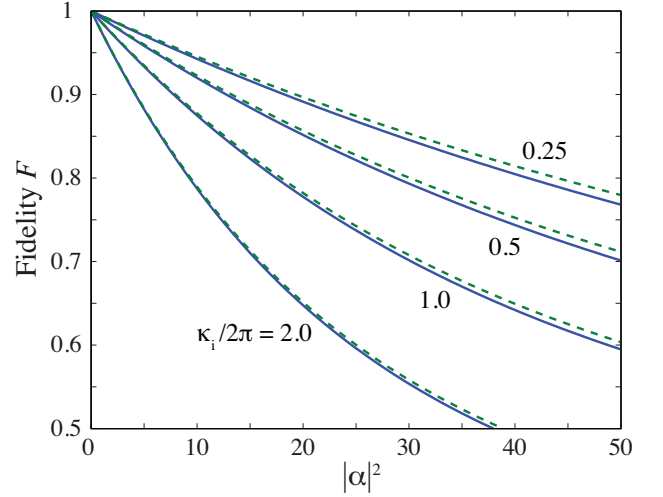}
  \caption{\label{fig:F_ki} (Color online). Fidelity, $F$, of entangled-path coherent state preparation as a function of the mean photon number, $|\alpha|^2$, in the incident pulse for $\gtw /2\pi =100$~MHz, with $\ki /2\pi =\{ 0.25,0.5,1.0,2.0\}$~MHz (curves are labelled by these values in units of MHz). Dashed lines are the approximate expression (\ref{eq:F_approx}). Other parameters are $\{\kex,h,\gamma\}/2\pi=\{50,0,5.2\}$~MHz, $t_{\rm p}=159$~ns ($\kappa t_{\rm p}\simeq 50$), and $\omega_{\rm A}=\omega_{\rm C}$.}
\end{figure}

We evaluate the fidelity, (\ref{eq:F}), numerically and present some results in Figs.~\ref{fig:F_g} and \ref{fig:F_ki}. In particular, we plot $F$ as a function of the mean photon number in the input pulse, $|\alpha|^2$, firstly in Fig.~\ref{fig:F_g} for several values of the atom-field coupling strength, $\gtw$, and secondly in Fig.~\ref{fig:F_ki} for a range of values of the intrinsic cavity mode loss rate, $\ki$. Considering Fig.~\ref{fig:F_g}, we see that an entangled coherent state with pulses of mean photon number $|\alpha|^2\simeq 10$ could in principle be achieved with fidelity in excess of 85\% for $\gtw /2\pi \gtrsim 50$~MHz and $\ki /2\pi \simeq 0.5$~MHz. With larger values of $\gtw$ and/or smaller values of $\ki$, similar fidelities appear possible with pulses of even larger mean photon numbers, as illustrated in Figs.~\ref{fig:F_g} and \ref{fig:F_ki}. 

\begin{figure}[t]
  \centering
  \includegraphics[scale=0.46]{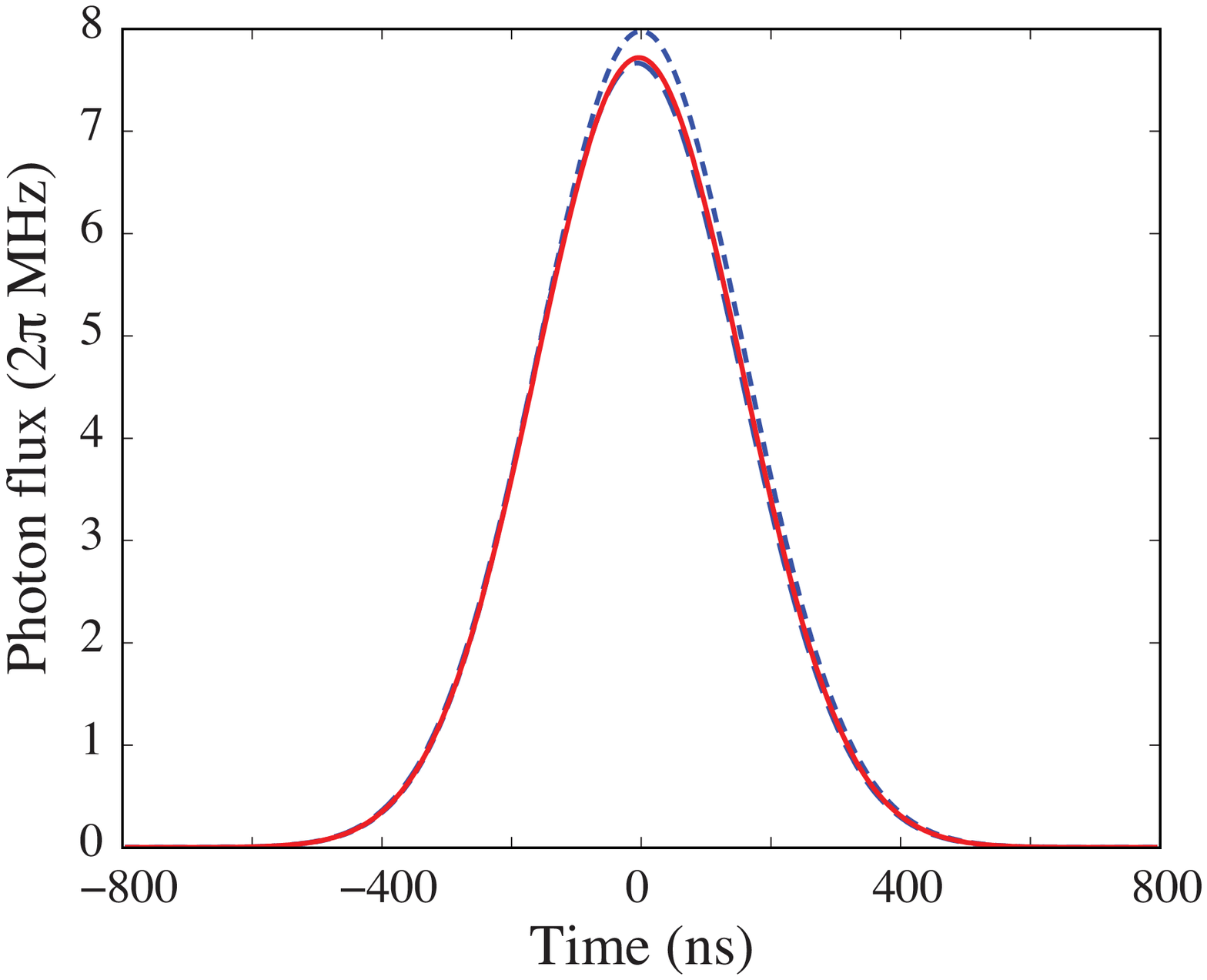}
  \caption{\label{fig:pulses} (Color online). Input photon flux, $|\braket{a_{\rm in,ex}(t)}|^2$ (blue, short-dashed curve), and output photon fluxes, $|\braket{a_{\rm out,ex}^{(0)}(t)}|^2$ (blue, long-dashed curve) and $|\braket{b_{\rm out,ex}(t)}|^2$ (red solid curve) as a function of time, for 
parameters $\{\gtw,\kex,\ki,h,\gamma\}/2\pi=\{100,50,0.5,0,5.2\}$~MHz, $\Ep /2\pi =28$~MHz, and $t_{\rm p}=159$~ns ($\kappa t_{\rm p}\simeq 50$), with $\omega_{\rm A}=\omega_{\rm C}$. The mean photon number in the input pulse (i.e., the area under the dashed curve) is $|\alpha|^2=20$. The output pulses have mean photon numbers $|\alpha_{\rm ex}^{(0)}|^2=19.2$ and $|\beta_{\rm ex}|^2=19.3$, respectively.}
\end{figure}

Implicit in the achievement of high fidelity state preparation is good matching, or overlap, between input and output pulses. An example of these pulses is shown in Fig.~\ref{fig:pulses}, where we plot the photon fluxes $|\braket{a_{\rm in,ex}(t)}|^2$, $|\braket{a_{\rm out,ex}^{(0)}(t)}|^2$, and $|\braket{b_{\rm out,ex}(t)}|^2$, where $a_{\rm out,ex}^{(0)}(t)$ denotes the output field for the atom in state $\ket{{\rm g}'}$. For the parameters used, the output pulses overlap extremely well with eachother and with the input pulse; the fidelity $F\simeq 0.85$ for this case. Note that on the scale used in Fig.~\ref{fig:pulses}, $|\braket{a_{\rm out,ex}(t)}|^2$ (not shown) is virtually indistinguishable from zero for all $t$ (and $|\alpha_{\rm ex}|^2=0.0017$).

Provided the bandwidth of the incident pulse is sufficiently narrow, i.e., $\kappa\tp\gg 1$, then the various integrals appearing in the expression for $F$ can be simplified; for example, we may set $|\alpha_{\rm ex}|^2 \simeq |t_{\rm ex}(0)|^2 \int_{-\infty}^\infty |\braket{\tilde{a}_{\rm in,ex}(\omega)}|^2 d\omega = |t_{\rm ex}(0)|^2|\alpha|^2$, and it is possible to derive the following approximate result for the fidelity,
\begin{align}
F & \simeq \frac{1}{4} \exp\left( -\Gamma_1 |\alpha|^2\right) + \frac{1}{4} \exp\left( -\Gamma_2 |\alpha|^2\right) \nonumber
\\
&\, +\, \frac{1}{2} \exp\left[ -\frac{1}{2}(\Gamma_1 +\Gamma_2 + 2\Gamma_3)|\alpha|^2 \right] ,
\label{eq:F_approx}
\end{align}
with
\begin{align}
\Gamma_1 &= \left( 1 - \frac{\kex}{\kappa}\frac{4C+2}{4C+1} \right)^2 + \left( 1 - \frac{\kex}{\kappa}\frac{4C}{4C+1} \right)^2 ,
\\
\Gamma_2 &= 4\left(\frac{\ki}{\kappa}\right)^2 ,
\\
\Gamma_3 &= \frac{\kex}{\kappa} \left( \frac{4C}{4C+1} \right)^2 \left( \frac{\ki}{\kappa} + \frac{1}{4C} \right) .
\end{align}
The result (\ref{eq:F_approx}) is also plotted in Figs.~\ref{fig:F_g} and \ref{fig:F_ki} and shows good agreement with the full expression for the chosen values of the pulse length $\tp$. 
Note that $\xi\simeq\exp (-\Gamma_3|\alpha |^2)$ is typically the key factor in determining $F$ for the entangled state preparation since, for $C\gg 1$, one has $\Gamma_1\simeq 2(\ki /\kappa)^2$ and $\Gamma_2\simeq 4(\ki /\kappa)^2$, while $\Gamma_3\simeq\kex\ki /\kappa^2\simeq\ki /\kappa\gg\Gamma_1,\Gamma_2$ for strong overcoupling ($\ki /\kappa\ll 1$). Hence, if $\Gamma_{1,2}|\alpha|^2\ll 1$ then one can write $F\simeq\frac{1}{2}(1+{\rm e}^{-\Gamma_3|\alpha|^2})$, which, e.g., agrees reasonably well with the curve for $\gtw/2\pi =150$~MHz in Fig.~\ref{fig:F_g}.

In comparison, the fidelity for merely ideal reflection of a coherent state pulse by an atom in state $\ket{\rm g}$,
\begin{align}
F_{\rm refl} = |\braket{0|\alpha_{\rm ex}}|^2 |\braket{-\alpha |\beta_{\rm ex}}|^2 ,
\end{align}
is (in the same limits) given by $\exp (-\Gamma_1|\alpha |^2)$. Taking $\Gamma_1\simeq 2(\ki /\kappa)^2$ and the parameters of Fig.~\ref{fig:F_g} ($\ki/\kappa=0.016$), this gives $F_{\rm refl}\gtrsim 0.97$ for $|\alpha|^2=50$. This highlights the relative fragility of the entangled-path coherent state and makes explicit the requirements on the operating regime of the system for high-fidelity preparation of such states; in particular, one requires $\ki/\kex\ll 1$ and $C=\gtw^2/\kappa\gamma\gg 1$. These conditions should indeed be achievable with microtoroid cavity QED.

\section{Conclusion}

To summarize, we have presented a scheme for a single-atom quantum switch for coherent light fields based upon microtoroid cavity QED in a regime not previously considered, at least not in this particular context. It complements recent schemes employing TM-polarized fields by also enabling, in principle, 100\% contrast in transmission and reflection of TE-polarized fields depending on the presence or absence of a strongly-coupled atom. The scheme can potentially work for quite strong incident fields, allowing, for example, control of a coherent state pulse of significant photon number and high-fidelity preparation of entangled-path coherent states.

\begin{acknowledgments}

SP acknowledges support from the International Central Networks Fund of the University of Auckland and from the Marsden Fund of the Royal Society of New Zealand. TA acknowledges support from JSPS KAKENHI Grant Number 26707022, MATSUO Foundation, and The Murata Science Foundation.

\end{acknowledgments}

\appendix
\section{Quantum Langevin equations and output fields: linearized analysis in frequency space}

In the linear regime, for which the atom can be treated as another harmonic oscillator, quantum Langevin equations for the mode operators in frequency space can be written (for $h=0$) as
\begin{widetext}
\begin{align}
-i\omega\tilde{a}(\omega) &= -(\kappa+i\Dc)\tilde{a}(\omega) -i\gtw^\ast\tilde{c}(\omega) + \sqrt{2\ki}\, \tilde{a}_{\rm in,i}(\omega) + \sqrt{2\kex}\, \tilde{a}_{\rm in,ex}(\omega) ,
\\
-i\omega\tilde{b}(\omega) &= -(\kappa+i\Dc)\tilde{b}(\omega) -i\gtw\tilde{c}(\omega) + \sqrt{2\ki}\, \tilde{b}_{\rm in,i}(\omega) + \sqrt{2\kex}\, \tilde{b}_{\rm in,ex}(\omega) ,
\\
-i\omega\tilde{c}(\omega) &= -(\gamma/2+i\Da)\tilde{c}(\omega) -i\gtw\tilde{a}(\omega) -i\gtw^\ast\tilde{b}(\omega) + \sqrt{\gamma}\, \tilde{c}_{\rm in}(\omega) 
\end{align}
\end{widetext}
where, for convenience, we have written $\sigma^-\equiv c$, $\{ a_{\rm in,ex}(t),a_{\rm in,i}(t),b_{\rm in,ex}(t),b_{\rm in,i}(t),c_{\rm in}(t)\}$ are (independent) input field operators, and the Fourier-transformed operators are defined by
\begin{align}
\tilde{o}(\omega) = \frac{1}{\sqrt{2\pi}} \int_{-\infty}^\infty {\rm e}^{-i\omega t} o(t)\, dt .
\end{align}
Input-output relations for the various channels take the forms
\begin{align}
\tilde{a}_{\rm out,ex}(\omega) &= -\tilde{a}_{\rm in,ex}(\omega) + \sqrt{2\kex}\, \tilde{a}(\omega) , 
\\
\tilde{a}_{\rm out,i}(\omega) &= -\tilde{a}_{\rm in,i}(\omega) + \sqrt{2\ki}\, \tilde{a}(\omega) , 
\\
\tilde{b}_{\rm out,ex}(\omega) &= -\tilde{b}_{\rm in,ex}(\omega) + \sqrt{2\kex}\, \tilde{b}(\omega) ,
\\
\tilde{b}_{\rm out,i}(\omega) &= -\tilde{b}_{\rm in,i}(\omega) + \sqrt{2\ki}\, \tilde{b}(\omega) , 
\\
\tilde{c}_{\rm out}(\omega) &= -\tilde{c}_{\rm in}(\omega) + \sqrt{\gamma}\, \tilde{c}(\omega) . 
\end{align}
We assume a coherent state input to channel $a_{\rm in,ex}$, of amplitude $\braket{a_{\rm in,ex}(t)}\longleftrightarrow \braket{\tilde{a}_{\rm in,ex}(\omega)}$, and vacuum inputs to all other channels. The mean coherent amplitudes in the output fields can then be derived as
\begin{align}
\braket{\tilde{a}_{\rm out,ex}(\omega)} &= t_{\rm ex}(\omega) \braket{\tilde{a}_{\rm in,ex}(\omega)} ,
\\
\braket{\tilde{a}_{\rm out,i}(\omega)} &= t_{\rm i}(\omega) \braket{\tilde{a}_{\rm in,ex}(\omega)} ,
\\
\braket{\tilde{b}_{\rm out,ex}(\omega)} &= r_{\rm ex}(\omega) \braket{\tilde{a}_{\rm in,ex}(\omega)} ,
\\
\braket{\tilde{b}_{\rm out,i}(\omega)} &= r_{\rm i}(\omega) \braket{\tilde{a}_{\rm in,ex}(\omega)} ,
\\
\braket{\tilde{c}_{\rm out}(\omega)} &= s(\omega) \braket{\tilde{a}_{\rm in,ex}(\omega)} ,
\end{align}
where the coefficients are given by
\begin{widetext}
\begin{align}
t_{\rm ex}(\omega) &= -1 + \frac{2\kex}{\kappa+i\Dc-i\omega}\, \frac{(\kappa+i\Dc-i\omega)(\gamma/2+i\Da-i\omega)+|\gtw|^2}{(\kappa+i\Dc-i\omega)(\gamma/2+i\Da-i\omega)+2|\gtw|^2} ,
\\
t_{\rm i}(\omega) &= \frac{2\sqrt{\ki\kex}}{\kappa+i\Dc-i\omega}\, \frac{(\kappa+i\Dc-i\omega)(\gamma/2+i\Da-i\omega)+|\gtw|^2}{(\kappa+i\Dc-i\omega)(\gamma/2+i\Da-i\omega)+2|\gtw|^2} ,
\\
r_{\rm ex}(\omega) &= - \frac{2\kex}{\kappa+i\Dc-i\omega}\,  \frac{\gtw^2}{(\kappa+i\Dc-i\omega)(\gamma/2+i\Da-i\omega)+2|\gtw|^2} ,
\\
r_{\rm i}(\omega) &= - \frac{2\sqrt{\ki\kex}}{\kappa+i\Dc-i\omega}\,  \frac{\gtw^2}{(\kappa+i\Dc-i\omega)(\gamma/2+i\Da-i\omega)+2|\gtw|^2} ,
\\
s(\omega) &= - \frac{i\gtw\sqrt{2\kex\gamma}}{(\kappa+i\Dc-i\omega)(\gamma/2+i\Da-i\omega)+2|\gtw|^2} .
\end{align}
The states of the output fields are then given by
\begin{align}
\ket{\alpha_{\rm ex}}_{a_{\rm out,ex}} &= \exp \left( -|\alpha_{\rm ex}|^2/2\right) \exp\left[ \int_{-\infty}^\infty \braket{a_{\rm out,ex}(t)} a_{\rm out,ex}^\dagger (t) dt\right] \ket{0}_{a_{\rm out,ex}} ,
\\
\ket{\alpha_{\rm i}}_{a_{\rm out,i}} &= \exp \left( -|\alpha_{\rm i}|^2/2\right) \exp\left[ \int_{-\infty}^\infty \braket{a_{\rm out,i}(t)} a_{\rm out,i}^\dagger (t) dt\right] \ket{0}_{a_{\rm out,i}} ,
\\
\ket{\beta_{\rm ex}}_{b_{\rm out,ex}} &= \exp \left( -|\beta_{\rm ex}|^2/2\right) \exp\left[ \int_{-\infty}^\infty \braket{b_{\rm out,ex}(t)} b_{\rm out,ex}^\dagger (t) dt\right] \ket{0}_{b_{\rm out,ex}} ,
\\
\ket{\beta_{\rm i}}_{b_{\rm out,i}} &= \exp \left( -|\beta_{\rm i}|^2/2\right) \exp\left[ \int_{-\infty}^\infty \braket{b_{\rm out,i}(t)} b_{\rm out,i}^\dagger (t) dt\right] \ket{0}_{b_{\rm out,i}} ,
\\
\ket{\eta}_{c_{\rm out}} &= \exp \left( -|\eta |^2/2\right) \exp\left[ \int_{-\infty}^\infty \braket{c_{\rm out}(t)} c_{\rm out}^\dagger (t) dt\right] \ket{0}_{c_{\rm out}} ,
\end{align}
\end{widetext}
with
\begin{align}
|\alpha_{\rm ex}|^2 &= \int_{-\infty}^\infty |t_{\rm ex}(\omega)|^2 |\braket{\tilde{a}_{\rm in,ex}(\omega)}|^2 d\omega ,
\\
|\alpha_{\rm i}|^2 &= \int_{-\infty}^\infty |t_{\rm i}(\omega)|^2 |\braket{\tilde{a}_{\rm in,ex}(\omega)}|^2 d\omega ,
\\
|\beta_{\rm ex}|^2 &= \int_{-\infty}^\infty |r_{\rm ex}(\omega)|^2 |\braket{\tilde{a}_{\rm in,ex}(\omega)}|^2 d\omega ,
\\
|\beta_{\rm i}|^2 &= \int_{-\infty}^\infty |r_{\rm i}(\omega)|^2 |\braket{\tilde{a}_{\rm in,ex}(\omega)}|^2 d\omega ,
\\
|\eta |^2 &= \int_{-\infty}^\infty |s(\omega)|^2 |\braket{\tilde{a}_{\rm in,ex}(\omega)}|^2 d\omega .
\end{align}




\begin{thebibliography}{25}
\bibitem{CohStateApprox}
Although our full quantum simulations in Section~IV show the output field $a_{\rm out,ex}$ (with atom) to in general be strongly bunched, and hence not strictly coherent, its amplitude for the parameters we consider here is extremely small, and its overlap with a (coherent) vacuum field is essentially equal to one.
\end{thebibliography}
\end{document}